\documentclass[aps,prd,twocolumn,showpacs,amsmath,amssymb,nofootinbib]{revtex4-1}
\usepackage{graphicx}
\usepackage{color}
\usepackage{times}
\usepackage{inputenc}
\usepackage{bm}
\usepackage{ulem}
\usepackage{multirow}
\usepackage{float}
\usepackage{url}
\usepackage{natbib}
\usepackage{mathrsfs}
\usepackage{physics}
\usepackage{comment}
\usepackage[table,xcdraw]{xcolor}
\usepackage{booktabs}
\usepackage{comment}
\usepackage{soul}
\usepackage[caption=false]{subfig}
\usepackage[colorlinks=true,citecolor=blue,urlcolor=blue,linkcolor=blue]{hyperref}

\usepackage{tikz,xcolor,hyperref}
\definecolor{lime}{HTML}{A6CE39}
\DeclareRobustCommand{\orcidicon}{%
	\begin{tikzpicture}
	\draw[lime, fill=lime] (0,0) 
	circle [radius=0.16] 
	node[white] {{\fontfamily{qag}\selectfont \tiny ID}};
	\draw[white, fill=white] (-0.0625,0.095) 
	circle [radius=0.007];
	\end{tikzpicture}
	\hspace{-2mm}
}
\foreach \x in {A, ..., Z}{%
	\expandafter\xdef\csname orcid\x\endcsname{\noexpand\href{https://orcid.org/\csname orcidauthor\x\endcsname}{\noexpand\orcidicon}}
}

\begin{document}
\title{Spacetime Curvature as a Probe of Exotic Core Phases in Neutron Stars within Modified Gravity}

\author{Sayantan Ghosh\orcidA{}}
\email{sayantanghosh1999@gmail.com}
\author{Bharat Kumar\orcidB{}}
\email{kumarbh@nitrkl.ac.in}
\author{Subhash Mahapatra\orcidC{}}
\email{mahapatrasub@nitrkl.ac.in}
\affiliation{\it Department of Physics and Astronomy, National Institute of Technology, Rourkela 769008, India}
\date{\today}
\begin{abstract}
In this study, we investigate the effect of Energy-Momentum Squared Gravity (EMSG) on the curvature of neutron stars (NSs) by using three relativistic mean-field (RMF) equations of state (EOSs) and three hadron–quark phase transition (HQPT) EOSs. Neutron stars, with their extreme densities and strong gravitational fields, provide an ideal laboratory for testing General Relativity (GR) in the high-curvature regime and for exploring possible deviations via modified gravity. EMSG extends GR by including nonlinear terms involving the energy-momentum tensor, characterized by a coupling parameter $\alpha$. We focus on the Kretschmann, Ricci, and Weyl curvature scalars, analyzing their dependence on baryon density and radial coordinate for varying values of $\alpha$. Our results indicate that EMSG can significantly alter the curvature profiles of neutron stars. In particular, the magnitude of both Weyl and Kretschmann scalars increases (decreases) for a positive (negative) EMSG parameter, with the former exhibiting a larger dependence. Similarly, the surface curvature (SC) is notably affected by $\alpha$. Interestingly, we further observe distinct discontinuities in the curvature profiles at hadron-quark phase transitions, especially in the soft and intermediate HQPT models. These signatures may provide observable imprints of exotic core phases in neutron stars.
\end{abstract}
\maketitle
\section{Introduction}
\label{intro}

Neutron stars (NSs) provide unique laboratories for testing gravity in regimes of extreme density and curvature \cite{Glendenning, shapiro, Curvature1, Curvature2, Curvature3}. With typical central densities exceeding nuclear saturation and gravitational potentials, NS interiors probe the strong-field, high-curvature domain far beyond solar-system tests \cite{Kopeikin2014}. 

In recent years, multi-messenger observations have begun to pin down NS structure \cite{Abbott_2017,doi:10.1126/science.aap9811, Abbott_2020, PhysRevX.9.011001, Fragione_2021,sgrb1,sgrb2}. For example, the NICER X-ray telescope has measured pulse profiles from PSR J0030+0451, yielding a canonical mass $M\simeq1.44^{+0.15}_{-0.14}\,M_\odot$ and radius $R\simeq13.0^{+1.2}_{-1.1}\,$km \cite{Miller_2019}. Meanwhile, the LIGO/Virgo detection of GW170817 – the first binary NS merger – has constrained the binary tidal deformability and common NS radius (roughly $R\sim10$-$13$~km) through the gravitational-wave phasing \cite{GW170817, PhysRevLett.121.091102, PhysRevX.9.011001}. These observations, combined with precise pulsar mass measurements, yield tight constraints on the neutron-star equation of state (EOS). At the same time, they open the door to testing gravity itself: NSs and their mergers are sensitive probes of any deviations from General Relativity (GR) in the strong-field regime \cite{Li:2011vx, PhysRevX.11.041050, Miller_2019, Ghosh2025}.

Despite its many successes, GR \cite{Einstein1915} is expected to break down at some scale (e.g., due to dark matter/energy puzzles or singularities) \cite{COPELAND, Riess_1998, Perlmutter_1999}, and alternative theories of gravity are actively explored \cite{CAPOZZIELLO2011167, OLMO20201, NOJIRI201159, NOJIRI20171, CQ, doiNOJIRI, lobo, Sotiriou, OLMO, OLMO20201, Clifton_2012}. In particular, unresolved issues, such as the cosmological constant problem, singularities in gravitational collapse, etc., motivate modifications of the Einstein-Hilbert action \cite{Ghosh2025}. Many extensions introduce new curvature terms (e.g.,\ $f(R)$ gravity \cite{Sotiriou, Astashenok_2015, PhysRevD.93.023501}) or couplings between matter and geometry (e.g.,\ $f(R,T)$ \cite{PhysRevD.84.024020, PhysRevD.97.104041} or $f(R,T_{\mu\nu}T^{\mu\nu})$ \cite{Katırcı2014, PhysRevD.94.044002} theories). One recent proposal is Energy-Momentum Squared Gravity (EMSG), in which the action contains an extra term proportional to the quadratic contraction $T_{\alpha\beta}T^{\alpha\beta}$ of the energy-momentum tensor \cite{PhysRevD.94.044002, AKARSU2023101305, universe10090339}. In this theory, the field equations coincide with GR in vacuum, but acquire extra, density-dependent terms inside matter \cite{PhysRevD.94.044002}. Early studies showed that EMSG can avoid the big-bang singularity (giving a cosmic bounce) and yields a viable cosmological history \cite{PhysRevD.94.044002, PhysRevD.98.024031}. Importantly, since deviations from GR arise only when $T_{\mu\nu}\neq0$, EMSG effects are expected to be most pronounced in the high-curvature cores of compact objects \cite{PhysRevD.98.024031, universe10090339}. Thus, NSs (and black holes (BHs)) are ideal environments to search for EMSG signatures.

A number of works have applied EMSG to compact stars. Nari \& Roshan \cite{PhysRevD.98.024031} derived the modified Tolman-Oppenheimer-Volkoff (TOV) equations in EMSG and solved them for a simple polytropic EOS. They found that the NS mass-radius relation can change substantially: for positive coupling $\eta$ (their notation for $T^2$ coupling), stars can be more massive than in GR with the same central pressure, potentially accommodating $2M_\odot$ pulsars with ordinary EOS \cite{PhysRevD.98.024031}. Similarly, Akarsu et al. \cite{EMSG_OAkarsu} solved the hydrostatic equations for realistic EOS and showed that EMSG-induced corrections become ``pronounced in the high-density cores of neutron stars''. Using observed mass–radius measurements, they constrained the EMSG coupling parameter to be extremely small (typically $|\alpha|\lesssim10^{-37}$ in appropriate units) and discussed implications for the hyperon puzzle and early-universe cosmology \cite{EMSG_OAkarsu}. 

More recently, Alam et al. \cite{EMSG_NAlam} performed an extensive survey of NS models in EMSG using a wide range of nuclear EOS. They demonstrated that positive EMSG coupling effectively stiffens the EOS at low densities and softens it at high densities, leading to smaller central pressures and larger maximum masses for given EOS \cite{EMSG_NAlam}. In these models the pressure, sound speed, and compactness profiles are altered relative to GR \cite{EMSG_NAlam}. Similarly, studies of radial stability indicate that EMSG stars generally remain stable (until reaching a turning point) for allowed parameter ranges. Tidal deformability, compactness, and oscillation properties have been explored by Ghosh \cite{Ghosh2025}, where universal relations between the fundamental ($f$-)mode frequency, tidal Love number, and stellar compactness have been established to probe EMSG effects. By fitting observationally inferred masses and radii, it was shown that EMSG changes the $f$-mode frequencies through its impact on compactness and deformability \cite{Ghosh2025}. In particular, different EOS types (stiff vs.\ soft) respond differently under EMSG modification, and the resulting sound-speed profiles have been examined to ensure causality. In short, existing EMSG studies have covered NS mass–radius relations \cite{PhysRevD.98.024031, EMSG_NAlam}, stability and oscillation modes \cite{Ghosh2025}, and constraints on the coupling from astrophysical data \cite{Ghosh2025}.

Parallel work has considered other compact systems. For instance, exact solutions for charged black holes in EMSG have been found, showing how the quadratic matter term modifies the Reissner–Nordstr\"{o}m geometry \cite{PhysRevD.94.044002}. Extensions with nonlinear electrodynamics have been studied to generate regular (nonsingular) black holes and ``bounces'' within EMSG \cite{universe10090339}. Quasinormal modes (ringdown) analyses in generalized EMSG (with nonminimal matter couplings) reveal changes in black-hole spectroscopy at high multipoles \cite{PhysRevD.101.064021}. These studies touch on black hole horizons and their perturbations, but a systematic exploration of black hole thermodynamics (entropy, temperature, etc.) in EMSG has not yet appeared. 

Despite this rich literature on EMSG stellar structure and oscillations, curvature invariants in EMSG neutron stars have not been investigated. In GR, one often includes invariants such as the Ricci scalar $\mathcal{R}$, the Kretschmann scalar $\mathcal{K}\equiv \sqrt{R_{\mu\nu\rho\sigma}R^{\mu\nu\rho\sigma}}$, and the Weyl invariant $\mathcal{W}\equiv \sqrt{C_{\mu\nu\rho\sigma}C^{\mu\nu\rho\sigma}}$ to characterize spacetime geometry independently of coordinates \cite{shakerin}. These invariants highlight key physical features: for example, the Kretschmann scalar in Schwarzschild geometry, given by $\mathcal {K}=\frac{4 \sqrt{3} M}{r^3}$, remains finite at the horizon ($r=2M$) but diverges at $r=0$, signaling the presence of a true curvature singularity \cite{shakerin}. Scalar invariants also form the basis for ideas such as Penrose’s Weyl-curvature hypothesis, which relates gravitational entropy to ratios of invariants. In a modified theory like EMSG, the extra matter-squared terms alter the field equations and thus the resulting spacetime curvature. One expects that the interior curvature profiles of an EMSG NS to differ from the GR case in a way that depends on the coupling. Studying $\mathcal{R}$, $\mathcal{K}$, and other invariants will reveal how the ``strength'' of gravity is redistributed by the new terms, which could have implications for tidal forces, geodesic motion near the star, and the approach to singularity or bounce.

Together, these curvature scalars serve as diagnostics of the gravitational field and internal geometry of compact stars \cite{Carroll, Curvature1, Curvature2}. For example, in GR studies it has been noted that $\mathcal{R}$ and the contraction of the Ricci tensor $\mathcal{F}$ ($=\sqrt{R_{\mu\nu}R^{\mu\nu}}$) drop to zero at the surface, so that $\mathcal{K}$ and $\mathcal{W}$ carry the dominant information there. Our goal is to compute these invariants for neutron star models in EMSG and compare them to the GR case. While previous work has examined EMSG in compact-star contexts - including the structure of neutron stars and hybrid stars - a detailed study of curvature invariants in EMSG compact stars remains unexplored \cite{EMSG_OAkarsu, EMSG_NAlam, PRETEL2023169440, Ghosh2025, Curvature3,Curvature2}. We will fill this gap by investigating the scalar curvatures of NS in EMSG for several representative EOSs. Specifically, we consider six equations of state: three pure hadronic (relativistic mean-field) models (NL3 \cite{NL3}, IOPB-I \cite{IOPB-I}, G3 \cite{G3}) and three hybrid hadron-quark models of varying stiffness (Stiff, Intermediate, and Soft phase-transition EOSs) \cite{HQPT}. These choices cover a range of plausible high-density behaviors and have been used in previous compact star studies. Indeed, as we see later on, our investigation suggests that not only curvature scalars vary nontrivially with baryon density and radial coordinate in NS, but also, depending on the EOS involved, they can capture signatures of exotic core phases of NSs. In particular, for EOSs that exhibit hadron-quark phase transitions, the scalar curvatures show a sharp jump and discontinuity at the transition densities.   

This paper is organized as follows: Sec.~\ref{sec:TF} presents the theoretical framework of EMSG. We derive the modified TOV equations and obtain
expressions for the curvature invariants in terms of the metric and matter variables. Sec.~\ref{sec:EOS} describes the equations of state used in our analysis, including their nuclear matter parameters and how the hadron–quark transition is implemented. Sec.~\ref{sec:R&D} contains our results and discussion. We solve the EMSG-TOV equations for each EOS, compute mass-radius relations, and evaluate the radial and density profiles of $\mathcal{K}$ and $\mathcal{W}$. We compare these to the GR predictions and highlight how the EMSG effects depend on the coupling parameter and EOS stiffness. We summarize our conclusions in Sec. \ref{sec:con}. Throughout this work, we use a metric signature $(-,+,+,+)$ and adopt geometrized units with $G = c = \hbar = 1$.

\section{Theoretical Framework}
\label{sec:TF}
\subsection{Energy-Momentum Squared Gravity (EMSG)}
\label{sec:EMSG}
In EMSG, the gravitational action is modified by an extra term quadratic in the matter fields. The action can be written as \cite{AkarsuNew}: 
\begin{align}
S=\int \left[\frac{1}{2\kappa}\mathcal{R}+f (\mathcal{L}_{\mathrm{m}}, g_{\mu\nu}T^{\mu\nu}, T_{\mu\nu}T^{\mu\nu})+ \mathcal{L}_{\mathrm{m}}\right]\sqrt{-g}\,\mathrm{d}^4x,
\label{action}
\end{align}
where, $f$ denotes any analytic function of matter-related scalars, such as the matter Lagrangian ($\mathcal{L}_{\mathrm{m}}$) \cite{Harko2010}, the trace ($g_{\mu\nu}T^{\mu\nu}$) \cite{Harko2011}, or the contraction ($T_{\mu\nu}T^{\mu\nu}$) \cite{Katırcı2014,PhysRevD.94.044002,PhysRevD.97.024011,PhysRevD.96.123517}. The Ricci scalar is denoted by $\mathcal{R}$, with gravitational coupling $\kappa =8\pi G$, with $G$ being Newton's constant. The strength of the EMSG correction is governed by the term $T_{\mu \nu }T^{\mu \nu }$, which is scaled by a real constant parameter $\alpha$.
\\
The energy-momentum tensor ($T_{\mu\nu}$) in terms of $\mathcal{L}_{\mathrm{m}}$, is defined as \cite{EMSG_OAkarsu,EMSG_NAlam}
\begin{align}  \label{tmunudef}
T_{\mu\nu}=-\frac{2}{\sqrt{-g}}\frac{\delta(\sqrt{-g}\mathcal{L}_{\mathrm{m}})}{\delta g^{\mu\nu}}=g_{\mu\nu}\mathcal{L}_{\mathrm{m}}-2\frac{\partial \mathcal{L}_{\mathrm{m}}}{\partial g^{\mu\nu}},
\end{align}
which depends only on the metric tensor components. 
Now we can write the total matter Lagrangian part as \cite{AkarsuNew},
\begin{align}
\mathcal{L}_{\mathrm{m}}^{tot}=\mathcal{L}_{\mathrm{m}}+f.
\end{align}
Putting it in Eq. \eqref{tmunudef}, we get the total $T_{\mu\nu}$ as
\begin{equation}
\begin{aligned}
\label{1tmunudef}
T^{tot}_{\mu\nu}&=-\frac{2}{\sqrt{-g}}\frac{\delta(\sqrt{-g}\mathcal{L}^{tot}_{\mathrm{m}})}{\delta g^{\mu\nu}}\\
&=-\frac{2}{\sqrt{-g}}\frac{\delta(\sqrt{-g}\mathcal{L}_{\mathrm{m}})}{\delta g^{\mu\nu}}-\frac{2}{\sqrt{-g}}\frac{\delta(\sqrt{-g}f)}{\delta g^{\mu\nu}}\,.
\end{aligned}
\end{equation}
Accordingly, we can write 
\begin{align}
\label{Ttot}
    T^{tot}_{\mu\nu} = T_{\mu\nu} + T^{mod}_{\mu\nu},
\end{align}
where 
\begin{align}
\label{tmod}
    T^{mod}_{\mu\nu}=-\frac{2}{\sqrt{-g}}\frac{\delta(\sqrt{-g}f)}{\delta g^{\mu\nu}}.
\end{align}
Similar to Eq. \eqref{tmunudef}, we can express Eq. \eqref{tmod} as \cite{AkarsuNew},
\begin{align}\label{2tmunu}
    T^{mod}_{\mu\nu}=-\frac{2}{\sqrt{-g}}\frac{\delta(\sqrt{-g}f)}{\delta g^{\mu\nu}} = fg_{\mu\nu}-2f_{T^2}\theta_{\mu\nu},
\end{align}
where 
\begin{align}\label{ftheta}
    f_{T^2}=\frac{\partial f}{\partial(T_{\rho\sigma}T^{\rho\sigma})},~~ 
    \theta_{\mu\nu}= \frac{\delta(T_{\rho\sigma}T^{\rho\sigma})}{\delta g^{\mu\nu}}.
\end{align}
In this study, we consider the function $f$ as 
\begin{align}
f(\mathcal{L}_{\mathrm{m}}, g_{\mu\nu}T^{\mu\nu}, T_{\mu\nu}T^{\mu\nu})=\alpha T_{\mu\nu}T^{\mu\nu},
\end{align}
which is the form of EMSG considered in \cite{Ghosh2025,Katırcı2014,PhysRevD.94.044002,PhysRevD.97.024011,PhysRevD.96.123517,EMSG_OAkarsu,EMSG_NAlam}.
Then, Eq.~\eqref{2tmunu} takes the following form
\begin{align}\label{3tmunu}
    T^{mod}_{\mu\nu}=\alpha T_{\rho\sigma}T^{\rho\sigma}g_{\mu\nu}-2\alpha \theta_{\mu\nu}.
\end{align}
Now, by using Eq. \eqref{Ttot}, we can write Einstein's field equations as
\begin{align} \label{Gmunu}
    G_{\mu\nu}=\kappa T_{\mu\nu} + \kappa T^{mod}_{\mu\nu}\,,
\end{align}
where $G_{\mu \nu }=\mathcal{R}_{\mu \nu }-\frac{1}{2}g_{\mu \nu }\mathcal{R}$ is the Einstein tensor. The ideal fluid form of $T_{\mu\nu}$ is given by
\begin{align}  \label{em}
T_{\mu\nu}=(\mathcal{E}+P)u_{\mu}u_{\nu}+P g_{\mu\nu},
\end{align}
where $\mathcal{E} $ and $P$ are the energy density and pressure respectively. $u_{\mu }$ is the four-velocity satisfying the conditions $u_{\mu }u^{\mu }=-1$, $\nabla _{\nu }u^{\mu }u_{\mu }=0$. Substituting into Eq.~\eqref{Gmunu}, the modified Einstein field equations in EMSG take the form
\begin{align}
G_{\mu\nu}=\kappa T_{\mu\nu}+\kappa \alpha \left(g_{\mu\nu}T_{\sigma\epsilon}T^{\sigma\epsilon}-2\theta_{\mu\nu}\right)\,,
\label{fieldeq}
\end{align}
where the new tensor $\theta _{\mu \nu }$ is defined as
\begin{equation}
\begin{aligned} \theta_{\mu\nu}=&~ T^{\sigma\epsilon}\frac{\delta
T_{\sigma\epsilon}}{\delta g^{\mu\nu}}+T_{\sigma\epsilon}\frac{\delta
T^{\sigma\epsilon}}{\delta g^{\mu\nu}} \\ =&-2\mathcal{L}_{\rm
m}\left(T_{\mu\nu}-\frac{1}{2}g_{\mu\nu}T\right)-TT_{\mu\nu} \\
&+2T_{\mu}^{\gamma}T_{\nu\gamma}-4T^{\sigma\epsilon}\frac{\partial^2
\mathcal{L}_{\rm m}}{\partial g^{\mu\nu} \partial g^{\sigma\epsilon}}.
\label{theta} \end{aligned}
\end{equation}
Here, $T = g^{\mu\nu}T_{\mu\nu}$ is the trace of $T_{\mu\nu}$. The final term of \eqref{theta} involves two derivatives of $\mathcal{L}_{\mathrm{m}}$, unlike Eq. \eqref{tmunudef}. For our cases of interest, this term is identically zero (see below); hence, we omit it from further analysis. A similar approach was adopted in \cite{AkarsuNew}.
Taking $\mathcal{L}_{\mathrm{m}}=P$ \cite{Faraoni,PhysRevD.109.104055}, we get the covariant divergence of Eq. \eqref{fieldeq} as
\begin{equation}
\nabla^{\mu}T_{\mu\nu}=-\alpha g_{\mu\nu}\nabla^{\mu}
(T_{\sigma\epsilon}T^{\sigma\epsilon})+2\alpha\nabla^{\mu}\theta_{\mu\nu}.
\label{nonconservedenergy}
\end{equation}
The local conservation of $T_{\mu\nu}$ holds only in the limit $\alpha =0$. Substituting Eq.~\eqref{em} into Eq.~\eqref{theta}, and then inserting the result into Eq. \eqref{fieldeq}, one gets
\begin{eqnarray}
& & G_{\mu\nu}=\kappa \mathcal{E} \left[\left(1+\frac{P}{\mathcal{E}}\right)u_{\mu}u_{\nu}+\frac{P}{\mathcal{E}}g_{\mu\nu}\right] \nonumber \\
&  &+\alpha\kappa\mathcal{E}^2\left[2\left(1+\frac{4P}{\mathcal{E}}+\frac{3P^2}{\mathcal{E}^2}\right)u_{\mu}u_{\nu}+\left(1+\frac{3P^2}{\mathcal{E}^2}\right)g_{\mu\nu}\right].\nonumber\\
\label{fieldeq2}
\end{eqnarray}
Now we can restore the Einstein field equation by redefining the above equation as
\begin{equation}
    G^{\mu\nu} = \kappa T^{\mu\nu}_{\mathrm{eff}},
    \label{EinsteinFEqn}
\end{equation}
where $T^{\mu\nu}_{\mathrm{eff}} = (\mathcal{E}_{\mathrm{eff}}+P_{\mathrm{eff}})u^{\mu}u^{\nu} + P_{\mathrm{eff}}g^{\mu\nu}$, is the effective energy-momentum tensor.
For an ideal fluid, $\mathcal{E}_{\mathrm{eff}}$ is the effective energy density defined as
\begin{equation}
\label{Eeff}
    \mathcal{E}_{\mathrm{eff}} = \mathcal{E} + \alpha\mathcal{E}^2\Bigg(1+\frac{8P}{\mathcal{E}} + \frac{3P^2}{\mathcal{E}^2}\Bigg)\,,
\end{equation}
and $P_{\mathrm{eff}}$ is the effective pressure defined as
\begin{equation}
\label{Peff}
    P_{\mathrm{eff}} = P + \alpha\mathcal{E}^2\Bigg(1+\frac{3P^2}{\mathcal{E}^2}\Bigg)\,.
\end{equation}
As noted in \cite{EMSG_OAkarsu}, the parameter $|\alpha |\sim\mathcal{E} ^{-1}$, and we know that for NSs, $\mathcal{E}\sim 10^{37}\,\mathrm{erg\,cm^{-3}}$ \cite{shapiro}. Consequently, EMSG corrections are expected to become important for compact objects such as NSs when the order of $\alpha$ is approximately $|\alpha |\sim 10^{-37}\,\mathrm{erg^{-1}\,cm^{3}}$.
\subsection{Hydrostatic equilibrium in EMSG}
\label{sec:hydrostatics}
\begin{figure*}
    \centering
    \includegraphics[width=0.47\textwidth]{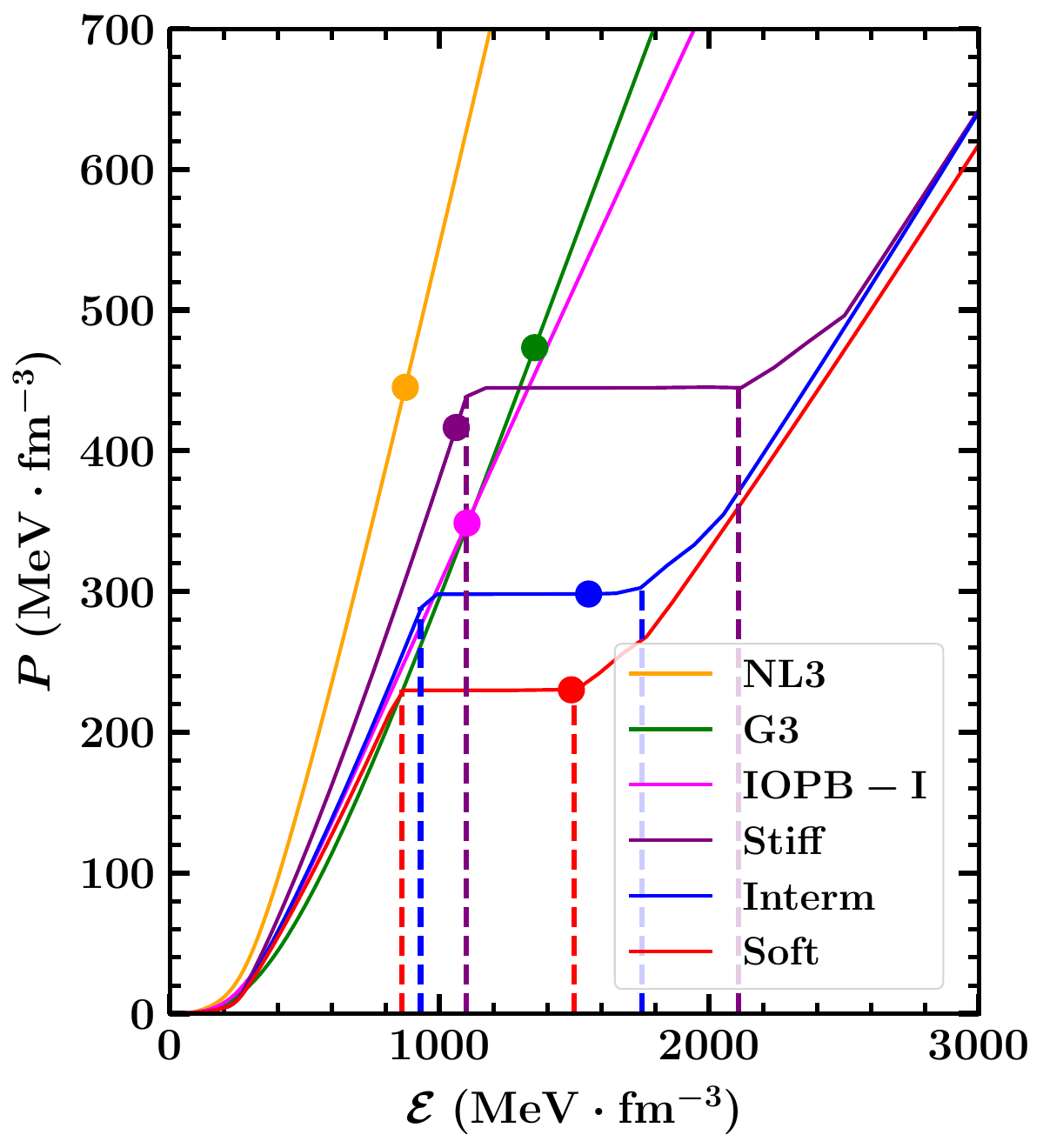}
    \includegraphics[width=0.47\textwidth]{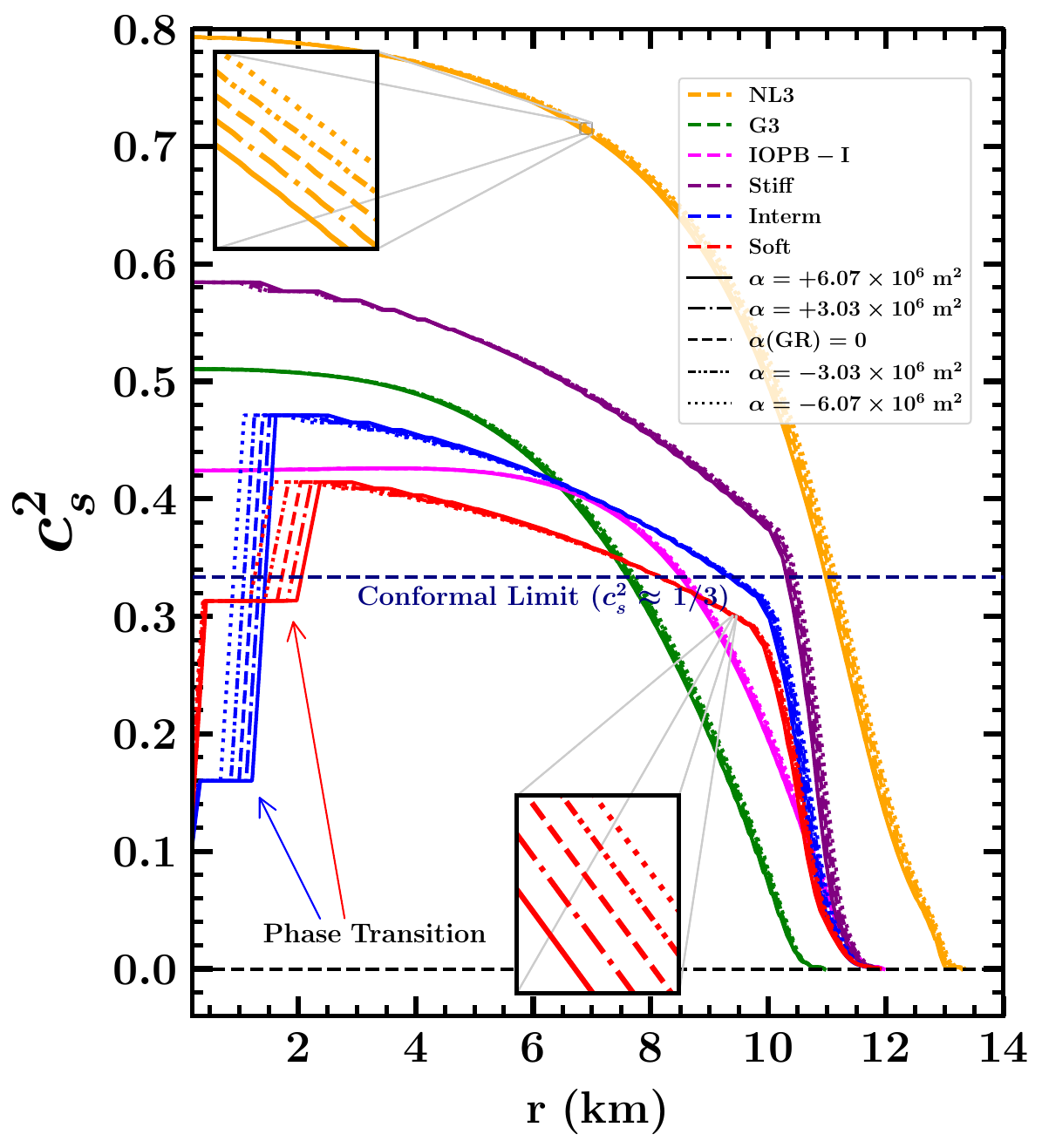}
    \caption{\textit{Left:} Variation of pressure ($P$) with energy density ($\mathcal{E}$) for NL3, G3, IOPB-I, Stiff, Intermediate, and Soft HQPT EOSs, represented by different colours. Bubble markers represent the energy density and pressure values corresponding to the maximum mass configuration for respective EOSs. \textit{Right:} The radial profile of sound speed ($c^2_s$) for alpha ($\alpha$) values of the maximum mass NS corresponding to respective EOSs. Zoom plot refers to the clear visibility of the effect of $\alpha$.}
    \label{EOS}
\end{figure*}

The spacetime geometry inside a static, spherically symmetric star is described by the line element \cite{Wald:1984rg, Schwarzschild:1916uq}:
\begin{equation}  
\label{eqn:metric}
\mathrm{d} s^2 = -e^{2\nu\left(r\right)}\mathrm{d} t^2 +e^{2\lambda\left(r\right)}\mathrm{d} r^2+r^2\mathrm{d}\theta^2+r^2\sin^2\theta \, \mathrm{d}\phi^2
\end{equation}
where $\nu(r)$ and $\lambda(r) $ are the metric functions. Substituting the metric into Eq. \eqref{fieldeq2}, we obtain the following set of independent equations:
\begin{equation}
\begin{aligned}
\frac{1}{r^2} - \frac{e^{-2\lambda}}{r^2} \left(1 - 2r \frac{\mathrm{d} \lambda}{\mathrm{d} r} \right) = \kappa \mathcal{E} + \kappa \alpha \mathcal{E}^2 \left(1 +  \frac{8P}{\mathcal{E}} +  \frac{3P^2}{\mathcal{E}^2} \right),
\end{aligned}
\label{eqn:f1}
\end{equation}
\begin{equation}
\begin{aligned}
-\frac{1}{r^2} + \frac{e^{-2\lambda}}{r^2} \left(1 + 2r \frac{\mathrm{d} \nu}{\mathrm{d} r} \right) = \kappa P + \kappa \alpha \mathcal{E}^2 \left(1 +  \frac{3P^2}{\mathcal{E}^2} \right).
\end{aligned}
\label{eqn:f2}
\end{equation}
Solving the modified Einstein equations in the EMSG framework leads to the modified TOV equations, given by
\begin{align}
\frac{\mathrm{d} m}{\mathrm{d} r}=4\pi r^2 \mathcal{E} \left[1+\alpha\mathcal{E} \left(1+\frac{8P}{\mathcal{E}}+\frac{3P^2}{\mathcal{E}^2 }\right)\right],  \label{TOV1}
\end{align}
\begin{align}
\frac{\mathrm{d} P}{\mathrm{d} r}&= -\frac{m\mathcal{E} }{r^2}\left(1+\frac{P}{\mathcal{E}}\right) \left( 1-\frac{2m}{r}\right)^{-1}  \notag \\
&\times \left[ 1+\frac{4\pi r^3 P}{m }+\alpha \frac{4\pi r^3\mathcal{E}^2}{m}\left(1+\frac{3P^2}{\mathcal{E}^2}\right)\right] \notag \\
&\times \left[1+2\alpha\mathcal{E}\left(1+\frac{3P}{\mathcal{E}} \right)\right] \left[1 + 2\alpha\mathcal{E} \left(c_s^{-2}+\frac{3P}{\mathcal{E}}\right)\right]^{-1}.  \label{TOV2}
\end{align}
We solved the TOV equations by integrating Eqs.~\eqref{TOV1}-\eqref{TOV2} from $r=0$, where $m(r=0)=0$ and $P(r=0)=P_c$ (the central pressure), upto the stellar surface $r=R$, where $m(r=R)=M$ and $P(r=R)=0$, by specifying a central energy density $\mathcal{E}(r=0)=\mathcal{E}_c$ at the center. This integration was performed for all six EOSs ($P(\mathcal{E})$) by varying the parameter $\alpha$ from negative to positive values to get the mass-radius profile. The profile $P(\mathcal{E})$ of these six EOSs is shown in the left panel of Fig.~\ref{EOS}. In this figure, the different color bubbles represent the energy density and pressure values corresponding to the maximum mass configuration, and the vertical dashed lines represent the phase transition regions for HQPT EOSs. For Soft and Intermediate HQPT EOSs, the bubbles are located within the phase transition region. However, for Stiff HQPT EOSs, the bubble (in purple color) is situated prior to (outside) the phase transition region; i.e., the maximum mass configuration is reached before the phase transition. Consequently, if we study the profile of any parameter of NSs for the maximum mass configuration, we would not see the phase transition effect on them for the Stiff HQPT EOS. We will see this behavior explicitly in the context of scalar curvatures in the subsequent discussion. Also, to validate the causality condition, we have calculated the radial profile of the sound speed for maximum mass configuration, defined as $c_{s}^2 \equiv \frac{\mathrm{d} P}{\mathrm{d} \mathcal{E}}$ (in units of $c^2$; $c$ is the speed of light). This is shown in the right panel of Fig.~\ref{EOS}. The causality limit is $0 \le c_{s}^2 \le 1$, and our calculated sound speed values are between $0$ to $0.8$. Accordingly, the causality condition is verified for all $\alpha$ variations. For the Soft (red) and Intermediate (blue) HQPT EOSs, the phase transition region is indicated by the arrows. As previously noted, for the Stiff (purple) EOSs, the maximum mass configuration occurs before the phase transition, so no visible effects of the phase transition are observed in $c_{s}^2$.
\subsection{Mathematical formulation for different Curvatures}
\label{sec:Curv}
In this section, we provide mathematical expressions of the scalar curvature quantities, mainly focusing on Kretschmann, full contraction of the Weyl and Ricci tensors, and Ricci scalars for a spherically symmetric NS in EMSG, using the effective $T_{\mu\nu}$ with $\mathcal{E} \equiv \mathcal{E}_{\mathrm{eff}}, P \equiv P_{\mathrm{eff}}$. We have adopted the mathematical form of these curvatures from \cite{Curvature1, Curvature2}.
\subsubsection{{Kretschmann Scalar ($\mathcal{K}$)}}
\label{Kret}
The Kretschmann scalar is defined as:
\begin{equation}
\mathcal{K}(r) \equiv \sqrt{R_{\mu\nu\rho\sigma} R^{\mu\nu\rho\sigma}},
\end{equation}
where $R_{\mu\nu\rho\sigma}$ is the Riemann curvature tensor.
After calculating the components of the Christoffel symbols and the Riemann tensor, applying the contraction, and using the modified TOV equations, we get the following expression of the Kretschmann scalar in EMSG:
\begin{equation}
\label{eqK}
\begin{split}
\mathcal{K}(r) = \Big[\, 
&(8\pi)^2 \left(3 \mathcal{E}_{\mathrm{eff}}^2 + 3 P_{\mathrm{eff}}^2 + 2 \mathcal{E}_{\mathrm{eff}} P_{\mathrm{eff}}\right) \\
&- \frac{128 \pi \mathcal{E}_{\mathrm{eff}} m(r)}{r^3}
+ \frac{48 m^2(r)}{r^6}
\,\Big]^{1/2}\,,
\end{split}
\end{equation}
where the effective energy density $\mathcal{E}_\mathrm{eff}$ and pressure $P_{\mathrm{eff}}$ are defined in Eqs.~\eqref{Eeff} and \eqref{Peff}, respectively.  $m(r)$ is the mass enclosed within the radius $r$.
Outside the star ($r > R$), where $P=0$, $\mathcal{E}=0$, and $m(r)=M$, the above expression simplifies to
\begin{equation}
\mathcal{K}(r) = \frac{4\sqrt{3} M}{r^3}.
\end{equation}
Similarly, on the surface of the star, we have
\begin{equation}
\label{Ksurface}
\mathcal{K}(R) = \frac{4 \sqrt{3} M}{R^3}.
\end{equation}
\subsubsection{{Weyl scalar  ($\mathcal{W}$)}}
\label{Weyl}
The Weyl tensor contraction is defined as:
\begin{equation}
\mathcal{W}(r) \equiv \sqrt{C_{\mu\nu\rho\sigma} C^{\mu\nu\rho\sigma}},
\end{equation}
where, $C_{\mu\nu\rho\sigma}$ is the Weyl tensor. The Weyl tensor is the traceless component of the Riemann tensor and is related to the Riemann tensor 
\begin{eqnarray}
&  & C_{\mu\nu\rho\sigma} = R_{\mu\nu\rho\sigma}  +\frac{R}{6}\left(g_{\mu\rho}g_{\nu\sigma} - g_{\mu\sigma}g_{\nu\rho}  \right) \nonumber \\ 
& &  +\frac{1}{2} \left(R_{\mu\sigma}g_{\nu\rho} -R_{\mu\rho}g_{\nu\sigma} + R_{\nu\rho}g_{\mu\sigma} -R_{\nu\sigma}g_{\mu\rho} \right)\,. 
\end{eqnarray}
The Weyl tensor $C_{\mu\nu\rho\sigma}$ shares the same symmetries as the Riemann tensor, but additionally satisfies the condition of being trace-free; i.e., metric contraction over any pair of indices vanishes. 

After calculating and contracting the components of the Weyl tensor for the spherical symmetry case in EMSG, we get the following expression:
\begin{align}
C_{\mu\nu\rho\sigma} C^{\mu\nu\rho\sigma}
&= \frac{4}{3} \left( \frac{6 m(r)}{r^3} - \kappa \mathcal{E}_{\mathrm{eff}} \right)^2 \\
&= \frac{4}{3} \left( \frac{6 m(r)}{r^3} - 8\pi \mathcal{E}_{\mathrm{eff}} \right)^2.
\end{align}
Accordingly, the full contraction of the Weyl tensor is
\begin{equation}
\label{eqW}
\mathcal{W}(r) = \left[ \frac{4}{3} \left( \frac{6 m(r)}{r^3} - 8\pi \mathcal{E}_{\mathrm{eff}}(r) \right)^2 \right]^{1/2}\,.
\end{equation}
Outside the star ($r > R$), the above expression reduces to
\begin{equation}
\mathcal{W}(r) = \frac{4 \sqrt{3} M}{r^3}.
\end{equation}
At the surface of the star, we have
\begin{equation}
\label{Wsurface}
\mathcal{W}(R) = \frac{4 \sqrt{3} M}{R^3}\,,
\end{equation}
Note that on the surface of the NS we have $\mathcal{W}(R)=\mathcal{K}(R)$.
\subsubsection{{Ricci Scalar ($\mathcal{R}$)}}
\label{Kret}
The Ricci scalar $\mathcal{R}$ is a scalar quantity derived from the Ricci curvature tensor, which in turn comes from the Riemann curvature tensor. It gives a measure of the curvature of spacetime due to mass-energy and is defined as 
\begin{equation}
\mathcal{R}=g_{\mu\nu}R^{\mu\nu}\,.
\end{equation}
The Ricci scalar provides a scalar measure of curvature that affects how volumes change in curved space. It can be simply calculated from the Einstein field equation by taking its trace. From Eq. \eqref{EinsteinFEqn}, we get
\begin{equation}
\mathcal{R}(r)=-\kappa T_{\mathrm{eff}}=\kappa\left(\mathcal{E}_{\mathrm{eff}}-3P_{\mathrm{eff}}   \right)\,.
\end{equation}
Note that, unlike the Kretschmann or Weyl scalar, the Ricci scalar goes to zero at or outside the surface of the star and has a nonzero contribution only in the interior of the star. 

\subsubsection{Surface Curvature (SC)}
Surface curvature (SC) is important to quantify the space-time wrap in the universe. It depends on the mass and the gravitational strength of the object. Since we are focusing on the Neutron star study, which is more dense and massive than the Sun, NSs should have more SC values. From \eqref{Ksurface} and \eqref{Wsurface}, we obtain the SC expressions corresponding to $\mathcal{K}$ and $\mathcal{W}$, which both yield the same values 
\begin{equation}
\mathcal{K}(R)=\mathcal{W}(R)= \mathrm{SC} = \frac{4 \sqrt{3} M}{R^3}\,,
\end{equation}
for a particular object of mass $M$ and radius $R$. Accordingly, the SC depends on the mass, radius, and  EOS corresponding to the composition within the star. For convenience, note that for the Sun, the value of the surface curvature is $\mathrm{SC}_\odot \approx 3.06 \times 10^{-23}m^{-2}$ \cite{Curvature1}.

\section{Equation of State}
\label{sec:EOS}
To calculate the curvature of the NS in EMSG, we choose three unified hadronic EOSs based on the RMF model and three HQPT EOSs.
\begin{enumerate}
\item {NL3 \cite{NL3}:}
Among the used RMF EOSs for this study, NL3 is the stiffest one, which includes nonlinear self-interaction terms for the $\sigma$ meson in the standard RMF model. It has a slope parameter $L=118.65\,\mathrm{MeV}$ and high incompressibility of $K=271.38\,\mathrm{MeV}$, which make it a stiff EOS. The maximum mass corresponding to this EOS is $M\approx2.774\, M_\odot$ and also has a large radius corresponding to the maximum mass. As a stiff EOS, its pressure varies stiffly with energy density, so it produces the lowest curvature values in comparison to the other two RMF EOSs used in this study.
\item {IOPB-I \cite{IOPB-I}:}
This EOS is developed within the framework of the extended relativistic mean field (E-RMF) model, incorporating the nonlinear vector and isoscalar-isovector coupling terms. The maximum mass corresponding to this EOS is $M \approx 2.149\,M_\odot$, which is consistent with current observational constraints. IOPB-I shows good agreement with the astrophysical data, including results from the GW170817 event \cite{GW170817}.
\item {G3 \cite{G3}:}
It is the softest EOS among the RMF EOSs used for this study, which is also an E-RMF parameterization that includes a set of additional coupling constants. The maximum mass corresponding to this EOS is $M \approx 1.997\, M_\odot$ and smaller radii in comparison to the NL3 and IOPB-I EOSs. Due to its soft nature, it has the highest curvature value.
\item {Stiff HQPT \cite{HQPT}:}
This is the stiffest EOS, among the HQPT EOSs used here, that's why named as Stiff EOS, which is constructed from the V-QCD model \cite{VQCD1, VQCD2}. The maximum mass corresponding to this EOS is $M \approx2.34\,M_\odot$, and has comparatively larger radii corresponding to the maximum mass.
\item {Intermediate (Interm) HQPT \cite{HQPT}:}
The intermediate HQPT EOS is also based on the V-QCD \cite{VQCD1, VQCD2} like the Stiff HQPT EOS, but has moderate stiffness. The maximum mass corresponding to this EOS is $M \approx2.14\,M_\odot$.
\item {Soft HQPT \cite{HQPT}:}
It is the softest HQPT EOS, which is constructed using the V-QCD \cite{VQCD1, VQCD2}, combined with the APR model \cite{APR} at low densities. Among the three HQPT EOSs, it has the lowest sound speed. Its softness leads to a lower maximum NS mass of $M \approx 2.02\,M_\odot$ with a smaller radius.
\end{enumerate}
The left panel of Fig.~\ref{EOS} shows the variation of pressure with energy density for the six unified EOSs mentioned above, with each EOS represented by a different color.
\begin{figure*}
    \centering
    \includegraphics[width=0.8\textwidth]{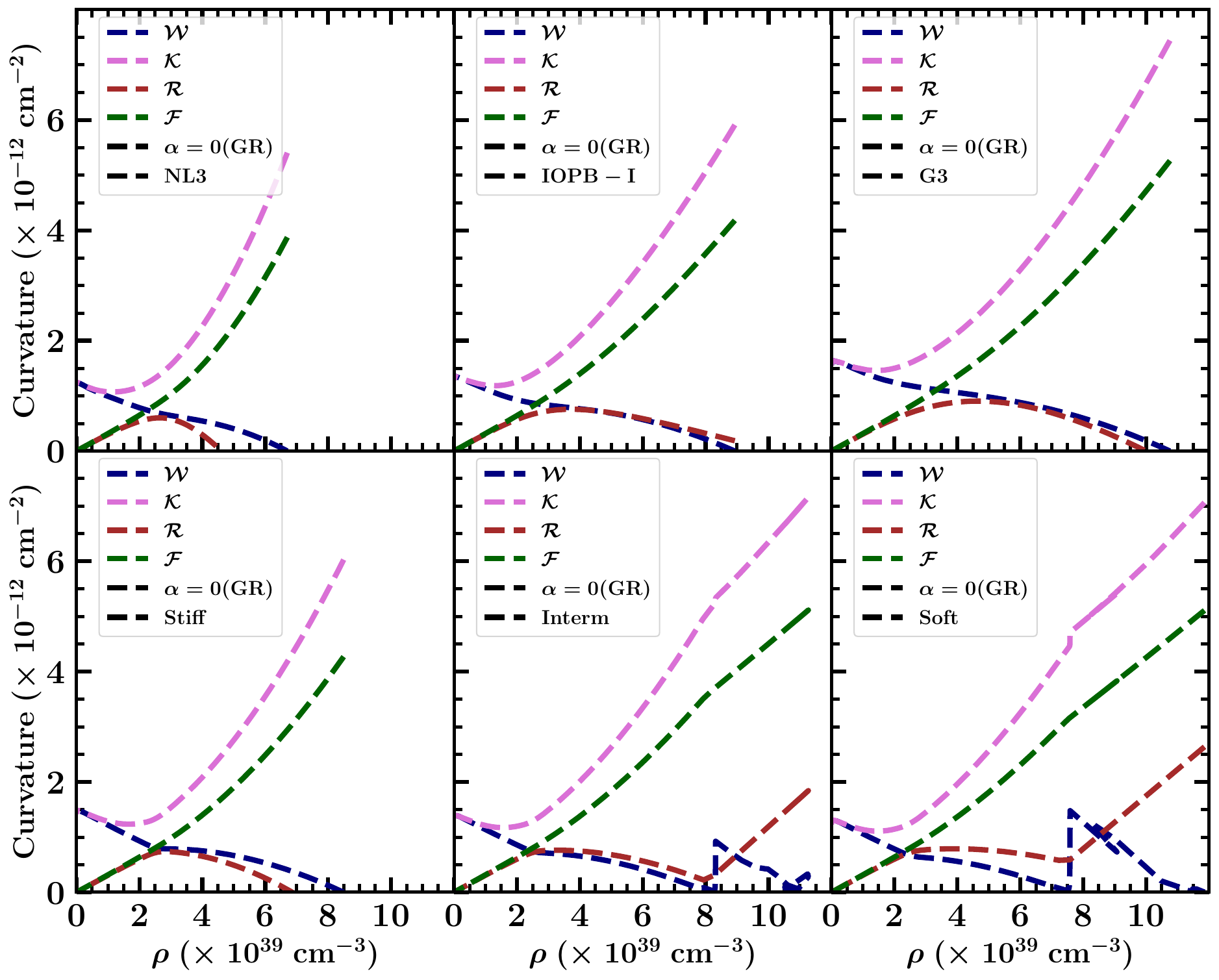}
    \caption{Variations of Ricci scalar ($\mathcal{R}$), full contraction of the Ricci tensor ($\mathcal{F}$), Kretschmann scalar ($\mathcal{K}$), and full contraction of Weyl tensor ($\mathcal{W}$) as functions of baryon density in GR ($\alpha = 0$).}
    \label{alpha0}
\end{figure*}
\section{Results and Discussion} 
\label{sec:R&D}
In this section, we examine the influence of the EMSG free parameter $\alpha$ on various spacetime curvature scalars across six representative EOSs: three based on RMF theory ({G3}, {IOPB-I}, and {NL3}) and three incorporating a hadron--quark phase transition (HQPT) ({Stiff}, {Intermediate}, and {Soft}). In all plots, $\alpha$ is expressed in units of $\mathrm{m}^2$.\footnote{{The original unit of $\alpha$ is $\mathrm{cm}^3/\mathrm{erg}$, which is converted to $\mathrm{m}^2$ for consistency with the geometric formalism. See Table~\ref{WKtable} for the numerical values used.}} 

Fig.~\ref{alpha0} shows the curvature invariants as functions of baryon density $\rho$ for all six EOSs in the pure GR case ($\alpha=0$).  In every case, $\mathcal{K}$ (pink) rises most steeply with density, dominating the core curvature.  The Ricci contraction $\mathcal{F}$ (green) grows monotonically but remains smaller than $\mathcal{K}$.  The Weyl scalar $\mathcal{W}$ (blue) generally decreases with $\rho$, reflecting that it approaches zero at the center.  The Ricci scalar $\mathcal{R}$ (red) is relatively small: it increases at low $\rho$, peaks at mid-density, and then declines.  As expected, toward the surface ($\rho\to0$), $\mathcal{R}$ and $\mathcal{F}$ vanish while $\mathcal{K}\approx\mathcal{W}$, consistent with the exterior Schwarzschild vacuum.  Across EOSs, the softest models reach the highest central densities and curvatures.  For instance, the G3 (soft RMF) and Soft HQPT EOS produce the largest central $\mathcal{K}$ and $\mathcal{F}$ values, whereas the stiff EOSs are less compact.  In the HQPT EOSs we observe nonmonotonic features at the transition density: small kinks in $\mathcal{R}$ and spikes in $\mathcal{K},\mathcal{W}$ appear around $\rho\sim7$–$9\times10^{39}\,\mathrm{cm}^{-3}$ for the Soft and Intermediate models.  The Stiff HQPT model, by contrast, shows no such features since its central density does not reach the transition.  In all cases $\mathcal{K}\gg\mathcal{F}\gg|\mathcal{W}|,|\mathcal{R}|$ in the core.  These GR-based results lay the foundation for assessing EMSG deviations in the following figures.
\begin{figure*}
    \centering
    \includegraphics[width=\textwidth]{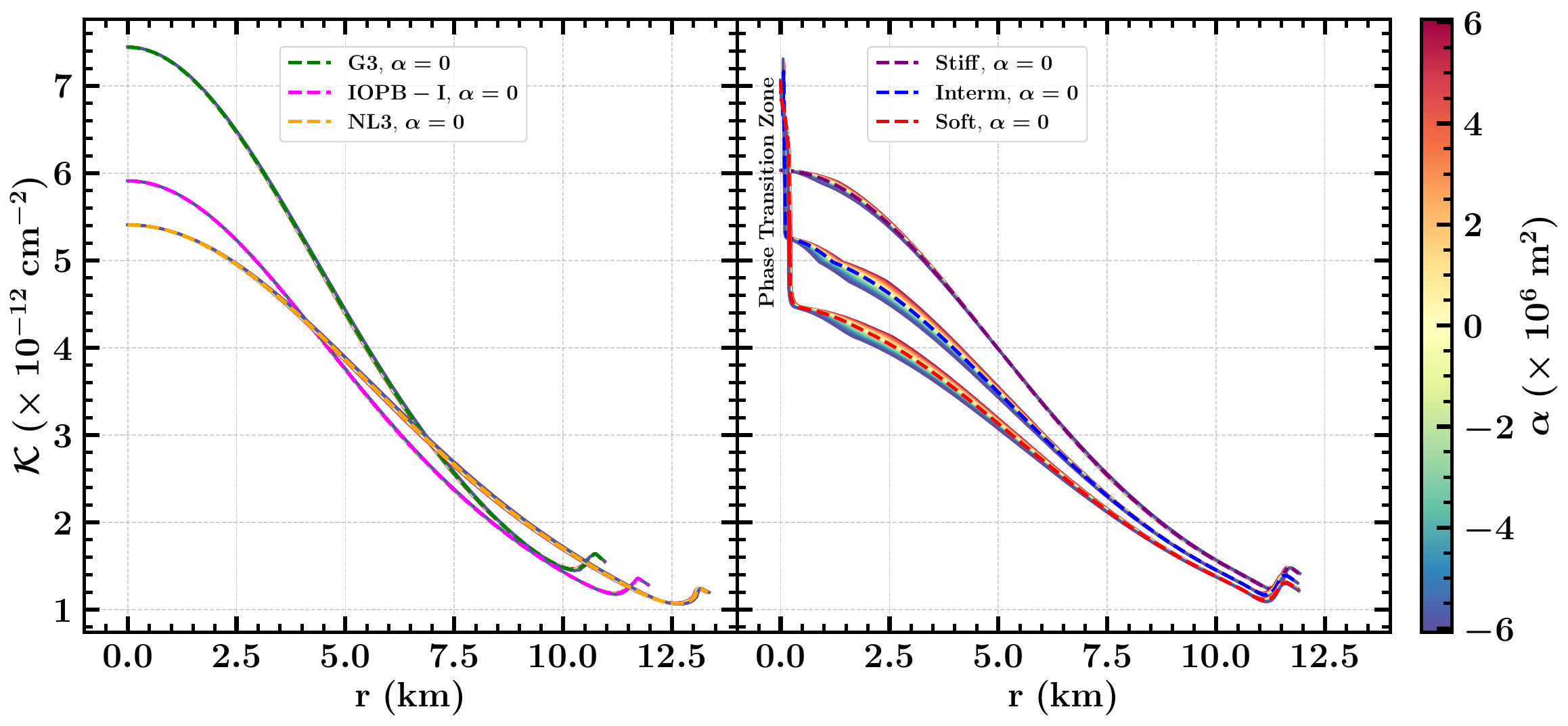}
    \caption{The radial variation of Kretschmann scalar curvature for three RMF (left panel) and three HQPT (right panel) EOSs. In the color bar, the variation of $\alpha$ has been shown.}
    \label{KR_RMF}
\end{figure*}
\begin{figure*}
    \centering
    \includegraphics[width=\textwidth]{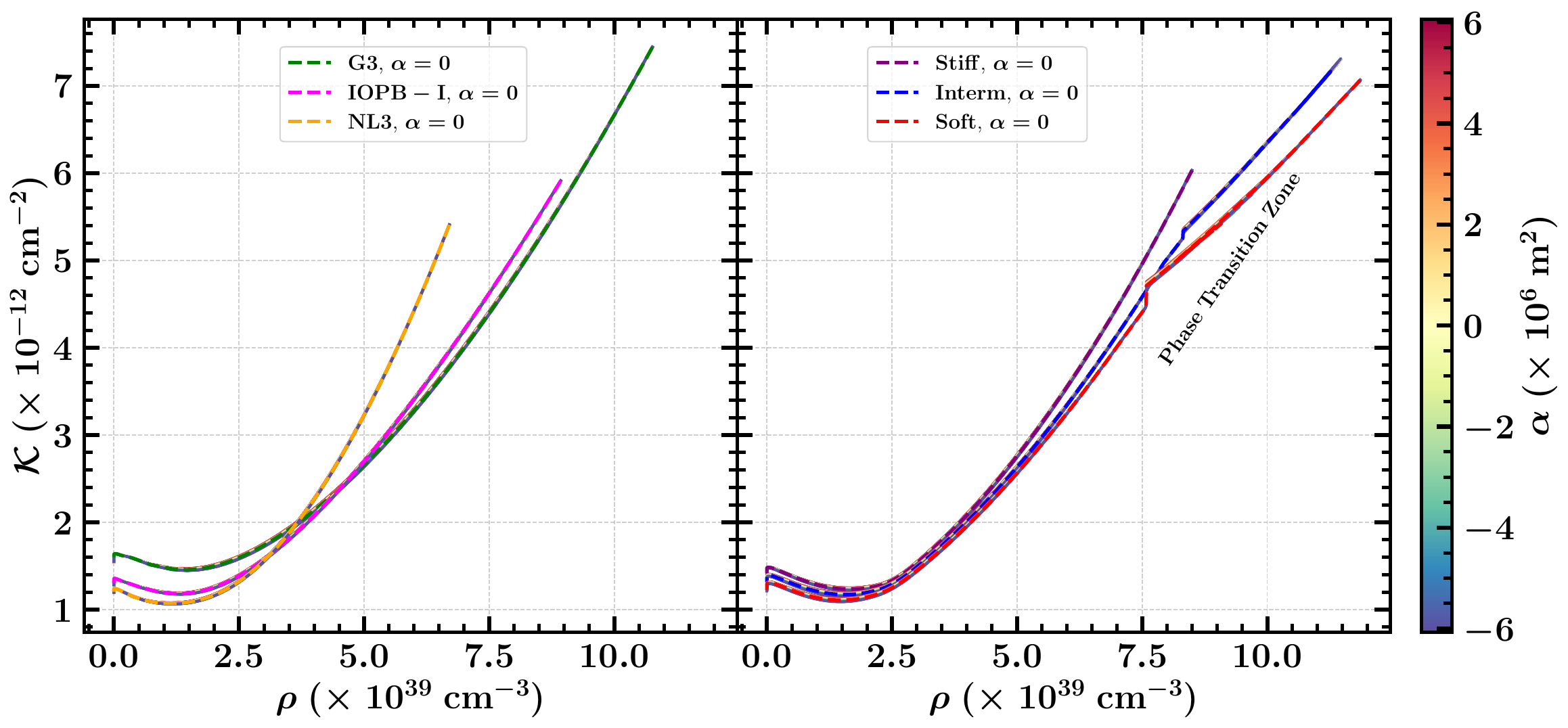}
    \caption{The variation of Kretschmann scalar curvature with baryon density ($\rho$) for three RMF (left panel) and three HQPT (right panel) EOSs. In the color bar, the variation of $\alpha$ has been shown.}
    \label{KRho_RMF}
\end{figure*}

In Fig.~\ref{KR_RMF}, the radial profiles of the Kretschmann scalar $\mathcal{K}(r)$ are presented for each EOS, with different values of $\alpha$ distinguished by color coding.  For pure hadronic stars (left panel), a stiffer EOS yields a larger radius and lower central curvature, whereas a softer EOS is more compact with higher $\mathcal{K}_{\rm center}$.  The central values occur at $r=0$ and $\mathcal{K}$ decreases monotonically outward.  In the HQPT models (right panel), the presence of a quark core produces a distinctive plateau: beyond the onset of quark matter, $\mathcal{K}$ remains roughly constant for a range of $r$ (the phase transition zone) before dropping near the surface.  This plateau is pronounced in the Soft and Intermediate EOSs (which undergo a transition) but absent in the Stiff EOS (which remains hadronic).  The color gradient shows the effect of $\alpha$: positive $\alpha$ increases $\mathcal{K}$ at fixed $r$ (making the star slightly more compact), while negative $\alpha$ reduces it.  These $\alpha$-induced changes are comparable in size to switching between different EOS stiffness.  

Fig.~\ref{KRho_RMF} shows $\mathcal{K}$ versus local density $\rho$ for the purely hadronic (left) and hybrid (right) cases.  As expected, away from the surface, as we move toward the core, $\mathcal{K}$ increases significantly with $\rho$ in all models, reflecting a stronger curvature in denser regions.  The RMF EOSs (left) yield smooth, continuous $\mathcal{K}(\rho)$ curves, with the softer EOS exhibiting a higher $\mathcal{K}$ value compared to the stiffer one near the core.  By contrast, in the HQPT EOSs (right), $\mathcal{K}$ undergoes an abrupt rise near the phase-transition region: the Soft and Intermediate curves exhibit a kink at $\rho\sim7$–$9\times10^{39}\,\mathrm{cm}^{-3}$ (the onset of quark deconfinement), while the Stiff EOS remains smooth.  The effect of $\alpha$ on these plots is most visible at high densities: at low $\rho$ all curves nearly coincide, but in the dense core region different $\alpha$ produce a spread.  Positive $\alpha$ shifts $\mathcal{K}(\rho)$ upward (softer effective core), whereas negative $\alpha$ shifts it downward.  This behavior is consistent with the fact that EMSG corrections become important only at extreme densities. Overall, as the value of $\alpha$ increases uniformly from a negative value to a positive value, $\mathcal{K}$, and hence the curvature of the NS, increases throughout the entire star.
\begin{figure*}
    \centering
    \includegraphics[width=\textwidth]{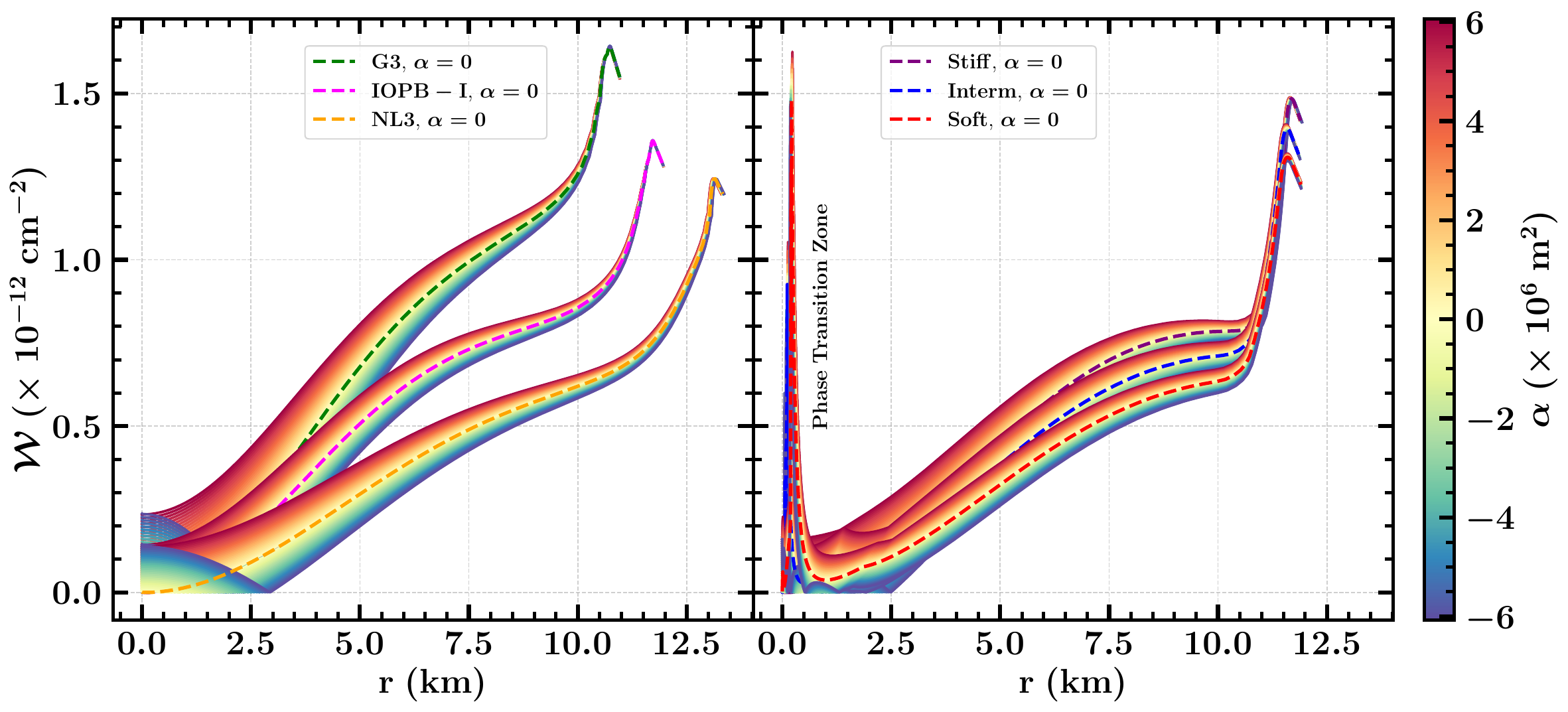}
    \caption{The radial variation of full contraction of the Weyl tensor curvature for three RMF (left panel) and three HQPT (right panel) EOSs. In the color bar, the variation of $\alpha$ has been shown.}
    \label{WR_RMF}
\end{figure*}
\begin{figure*}
    \centering
    \includegraphics[width=\textwidth]{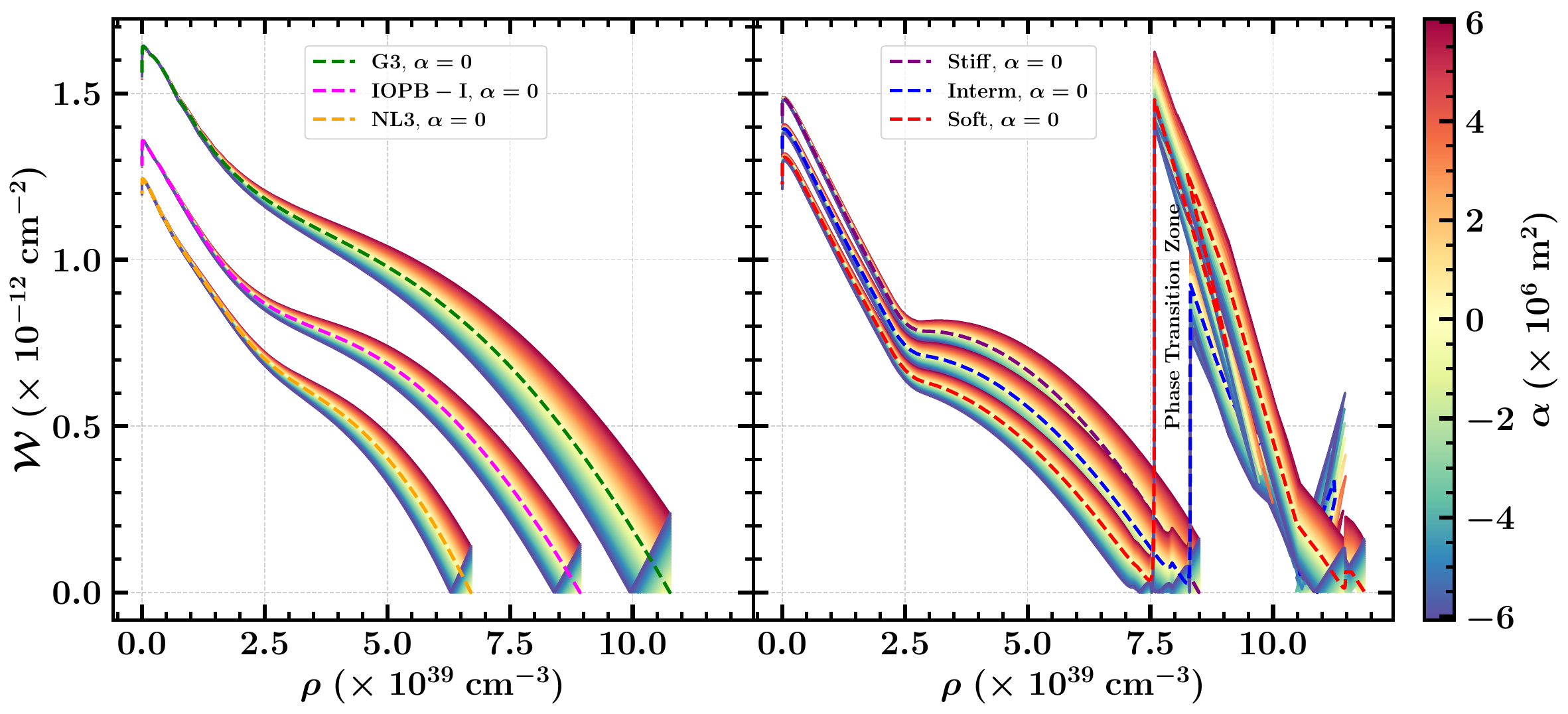}
    \caption{The variation of full contraction of the Weyl tensor curvature with baryon density ($\rho$) for three RMF (left panel) and three HQPT (right panel) EOSs. In the color bar, the variation of $\alpha$ has been shown.}
    \label{WRho_RMF}
\end{figure*}

Next, Fig.~\ref{WR_RMF} shows the radial profile of the Weyl scalar $\mathcal{W}(r)$.  Unlike $\mathcal{K}$, $\mathcal{W}$ vanishes at $r=0$ and rises outward to a maximum near the surface.  This reflects that tidal (trace-free) curvature accumulates away from the center. In the left panel (RMF EOSs), all $\mathcal{W}$ curves start at zero, increase to a peak, and then drop until the surface occurs at $r=R$.  Softer EOSs (like G3) produce larger interior gradients.  Notably, $\mathcal{W}$ shows stronger sensitivity to $\alpha$ than $\mathcal{K}$ does: the color bands (varying $\alpha$) are broader, especially in the inner core.  In the right panel (HQPT EOSs), the Soft and Intermediate models exhibit a sharp change in the $\mathcal{W}$ profile near the phase transition radius, analogous to the feature in $\mathcal{K}$.  The Stiff EOS again shows a smooth profile.  Overall, softer EOSs yield larger differences in $\mathcal{W}$ when $\alpha$ varies, indicating enhanced tidal deviations from GR.  

Fig.~\ref{WRho_RMF} shows the structure of $\mathcal{W}$ with $\rho$.  In all RMF EOSs based cases, $\mathcal{W}$ exhibits a peak when the density is small and then decreases monotonically with it (approaching zero at high $\rho$), i.e., the tidal curvature vanishes in the conformally-flat core.  The HQPT EOSs (right) again show kink structure at the transition: Soft and Intermediate have an abrupt slope change in $\mathcal{W}(\rho)$ around the same $\rho$ range as in Fig.~\ref{KRho_RMF}, signaling the phase transition.  The Stiff EOS remains smooth.  The spread of the colored bands (varying $\alpha$) is wider than for $\mathcal{K}$, especially for softer EOSs.  This confirms that $\mathcal{W}$ is more sensitive to EMSG coupling than $\mathcal{K}$. In particular, even a small $\alpha$ produces a noticeable shift in $\mathcal{W}$ at high densities, whereas $\mathcal{K}$ remains comparatively unchanged for the same $\alpha$-range. It is also straightforward to show that the difference $|\mathcal{K}-\mathcal{W}|$ is always positive for each value of $\alpha$, justifying our above discussion that the Kretschmann invariant is always higher than the Weyl invariant everywhere inside the NS. However, on the surface, they reduce to the same value, as can also be observed from Fig.~\ref{alpha0}.
\begin{figure*}
    \centering
    \includegraphics[width=\textwidth]{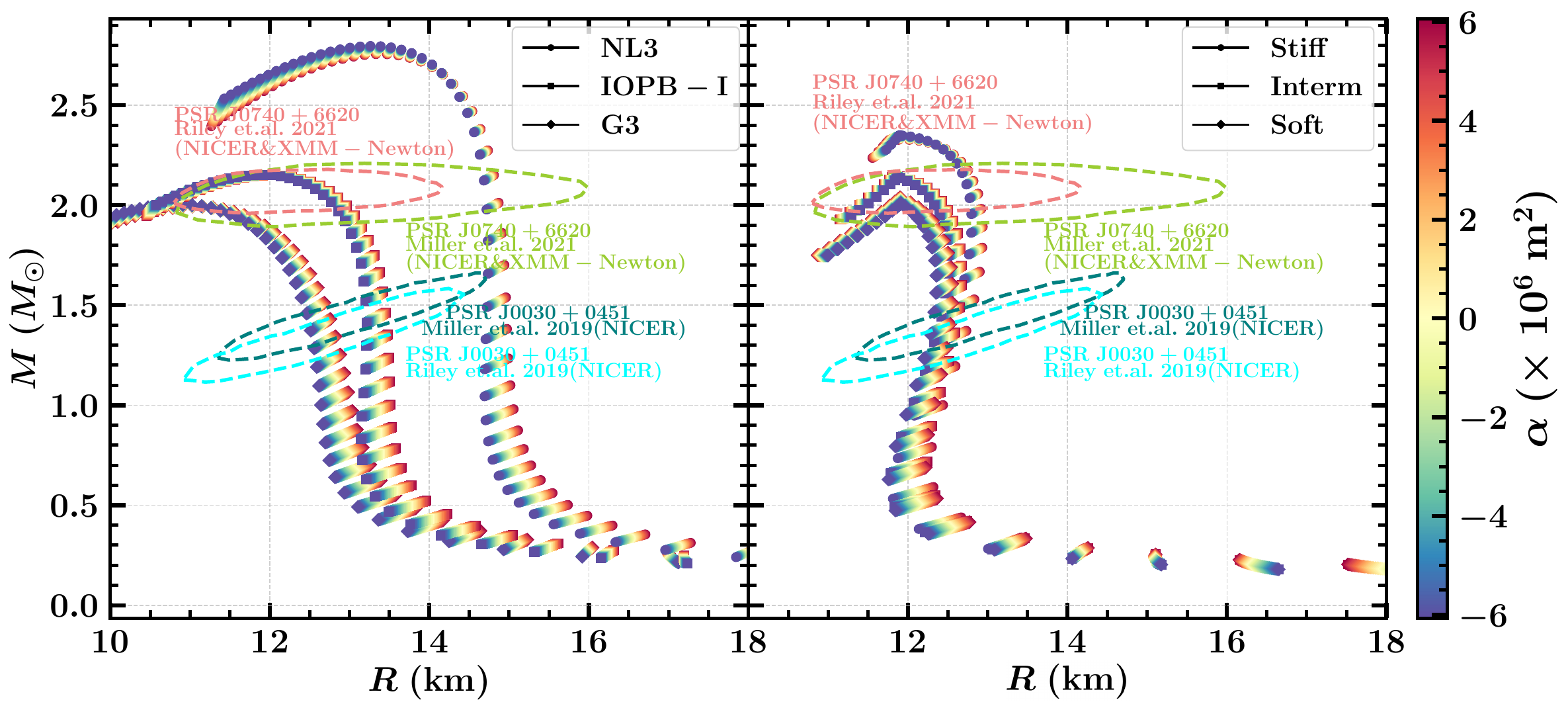}
    \caption{The Mass-Radius relationship obtained for our choice of EOSs, with the $\alpha$ parameter variation. The astrophysical observable constraints on mass and radius from PSR J0740+6620 \cite{Miller_2021, Riley_2021}, and NICER data for PSR J0030+0451 \cite{Miller_2019, Riley_2019} are represented by colored regions.}
    \label{MR}
\end{figure*}
\begin{figure*}
    \centering
    \includegraphics[width=\textwidth]{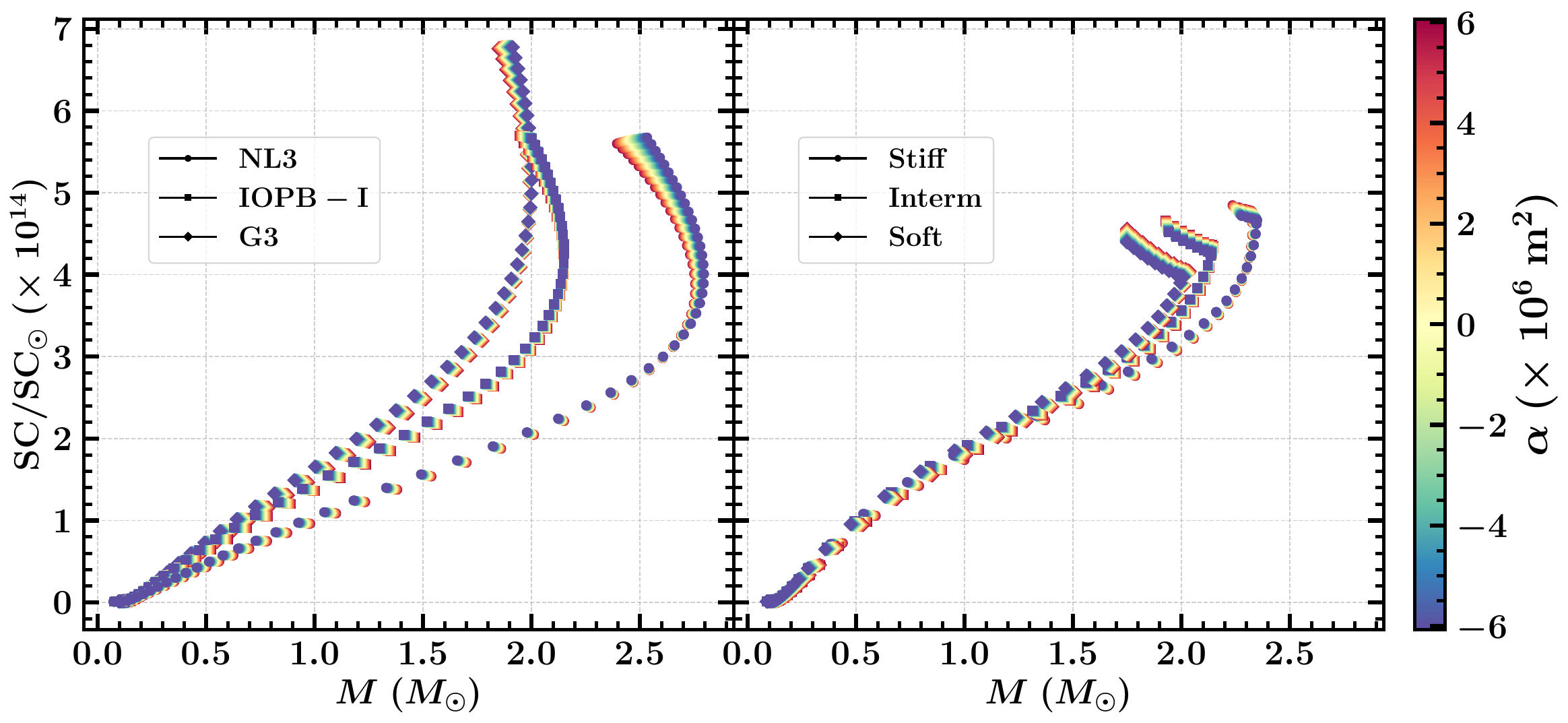}
    \caption{Variation of the surface curvature with the NS mass, with varying the $\alpha$ parameter. $\mathrm{SC}_\odot (\approx 3.06 \times 10^{-23} m^{-2})$ \cite{Curvature1} is the surface curvature for the Sun. The zoom plot shows the clear visibility of the effect of $\alpha$.}
    \label{SC}
\end{figure*}
\begin{table*}
\caption{Maximum mass ($M$), corresponding radius ($R$) and radius at 1.4M$_\odot$ ($R_{1.4}$); surface curvature corresponding to the maximum mass ($\mathrm{SC}$) and at 1.4M$_\odot$ ($\mathrm{SC}_{1.4}$) of the NS for the EOSs corresponding to {G3}, {IOPB-I}, {NL3}, {Stiff}, {Intermediate}, {Soft}, calculated with the variation of $\alpha$ parameters. $\mathrm{SC}_\odot \approx 3.06 \times 10^{-23}m^{-2}$; surface curvature for Sun.}
\centering
\setlength{\tabcolsep}{10pt}
\renewcommand{\arraystretch}{1}
\scalebox{1}{\begin{tabular}{ccccccc}
\hline
\hline\noalign{\smallskip} 
{\begin{tabular}[c]{@{}l@{}}EOS models\end{tabular}}&
{\begin{tabular}[c]{@{}l@{}}$\alpha$ $(10^{6}$ m$^2$)\end{tabular}}&
{\begin{tabular}[c]{@{}l@{}}$M$ ($M_\odot$)\end{tabular}}&
{\begin{tabular}[c]{@{}l@{}}$R$
(Km)\end{tabular}}&
{\begin{tabular}[c]{@{}l@{}}$R_{1.4}$ (Km)\end{tabular}}&
{\begin{tabular}[c]{@{}l@{}}$\mathrm{SC}$
$(10^{14}\mathrm{SC}_\odot)$  \end{tabular}}&
{\begin{tabular}[c]{@{}l@{}}$\mathrm{SC}_{1.4}$ $(10^{14}\mathrm{SC}_\odot)$ \end{tabular}}
\\
\hline
\hline\noalign{\smallskip} 
& $-6.07$   & 2.003 & 10.897 & 12.493  & 5.176 & 2.401    \\
& $-3.03$   & 2.000& 10.893 & 12.567   & 5.173 & 2.359     \\
{G3} & 0    & 1.997& 10.997 & 12.638   & 5.020 & 2.320   \\ 
& $+3.03$   & 1.995& 10.997 & 12.710   & 5.015 & 2.280     \\
& $+6.07$   & 1.993& 11.109 & 12.785   & 4.860 & 2.240    \\ 
\hline\noalign{\smallskip}  
& $-6.07$   & 2.151& 11.858 & 13.210  & 4.313 & 2.031   \\
& $-3.03$   & 2.150& 11.954 & 13.274  & 4.209 & 2.001     \\
{IOPB-I} & 0  & 2.150& 11.958 & 13.341   & 4.204 & 1.971   \\ 
& $+3.03$   & 2.149& 11.958 & 13.408  & 4.203 & 1.942    \\
& $+6.07$   & 2.149& 12.066 & 13.476  & 4.091 & 1.913    \\ 
\hline\noalign{\smallskip} 
& $-6.07$   & 2.797& 13.262 & 14.717  & 4.009 & 1.468    \\
& $-3.03$   & 2.785& 13.358 & 14.781  & 3.907 & 1.449     \\
{NL3} & 0   & 2.775& 13.342 & 14.847  & 3.906 & 1.430   \\ 
& $+3.03$   & 2.764& 13.326 & 14.914  & 3.906 & 1.411    \\
& $+6.07$   & 2.755& 13.434 & 14.981  & 3.799 & 1.392   \\ 
\hline\noalign{\smallskip} 
& $-6.07$   & 2.349& 11.898 & 12.537  & 4.663 & 2.376    \\
& $-3.03$   & 2.343& 11.874 & 12.587  & 4.680 & 2.345    \\
{Stiff} & 0 & 2.337& 11.842 & 12.643  & 4.706 & 2.316  \\ 
& $+3.03$   & 2.332& 11.974 & 12.696  & 4.542 & 2.286     \\
& $+6.07$   & 2.327& 11.970 & 12.750  & 4.537 & 2.257  \\ 
\hline\noalign{\smallskip} 
& $-6.07$   & 2.138& 11.886 & 12.386  & 4.258 & 2.462     \\
& $-3.03$   & 2.142& 11.870 & 12.442  & 4.282 & 2.430   \\
{Intermediate} & 0 & 2.145& 11.854 & 12.499  & 4.306 & 2.397    \\ 
& $+3.03$   & 2.147& 11.842 & 12.555  & 4.324 & 2.365     \\
& $+6.07$   & 2.149& 11.826 & 12.613  & 4.345 & 2.332   \\ 
\hline\noalign{\smallskip} 
& $-6.07$   & 2.006& 11.910 & 12.297  & 3.970 & 2.517     \\
& $-3.03$   & 2.015& 11.910 & 12.357  & 3.987 & 2.480     \\
{Soft} & 0  & 2.023& 11.906 & 12.416  & 4.008 & 2.445    \\ 
& $+3.03$   & 2.031& 11.906 & 12.475  & 4.023 & 2.411    \\
& $+6.07$   & 2.038& 11.902 & 12.531  & 4.041 & 2.379     \\ 
\noalign{\smallskip}\hline
\hline
\end{tabular}}
\label{WKtable}
\end{table*}

In Fig.~\ref{MR}, we present the mass-radius ($M$-$R$) relationships for neutron star modelled with six EOSs under variation of the EMSG parameter $\alpha$ (in units of $\mathrm{m}^2$; see Table~\ref{WKtable} for numerical values). Observational constraints from NICER and XMM-Newton measurements of PSR J0740+6620 and PSR J0030+0451 are overlaid for comparison, providing critical compatibility checks across the parameter space. The left panel shows results for three RMF EOSs: G3, IOPB-I, and NL3. Among these, NL3 is the stiffest, yielding the largest radii for a given mass. Across all three EOSs, the effect of the EMSG coupling parameter $\alpha$ exhibits a consistent pattern. 
Before reaching the maximum mass, the positive values of $\alpha$ increase the mass as well as the radius (see the fifth column ($R_{1.4}$ values) in Table~\ref{WKtable}). However, at the maximum mass and beyond, negative values of $\alpha$ increase the mass and decrease the stellar radius, effectively stiffening the equation of state. Conversely, positive $\alpha$ values reduce the mass and increase the radius, corresponding to an effective softening of the EOS. The right panel displays results for the three HQPT EOSs: Stiff, Intermediate, and Soft. Here also, we can see the same effect of $\alpha$ on mass and radius before reaching the maximum mass, like the figure in the left panel. However, at the maximum mass, the Stiff EOSs again exhibit trends qualitatively similar to those of the RMF models, while the Soft and Intermediate EOSs behave differently due to the phase transition. This deviation arises from the suppression or delayed onset of the quark phase at high central densities, which is sensitive to the value of $\alpha$.  

Finally, Fig.~\ref{SC} plots the surface curvature ${\rm SC}\equiv 4\sqrt{3}M/R^3$ (in units of the solar value ${\rm SC}_\odot$) as a function of the NS mass $M$. A clear $\alpha$-dependence emerges for all six EOSs. In particular, at fixed $M$, below the maximum mass, positive $\alpha$ reduces SC (since $R$ is slightly larger) and negative $\alpha$ increases SC.  The NS surface curvature is $\sim 10^{14}$ times the solar value, underscoring the extreme gravity.  We tabulate the maximum mass $M$, corresponding radius $R$, and SC (both at $M$ and at $M=1.4M_\odot$) in Table~\ref{WKtable} for each EOS and $\alpha$.  The results confirm that all stars remain causal ($0 \le c_{s}^2 \le 1$) and above $2M_\odot$ for $|\alpha|\le6\times10^6\,\mathrm{m}^2$.  
\section{Conclusion}
\label{sec:con}
In this work, we examined the curvature structure of neutron stars within the framework of Energy-Momentum Squared Gravity (EMSG), using six realistic EOSs (three hadronic RMF models and three hybrid HQPT models). Due to its significant implications in cosmology, EMSG has garnered increasing attention in recent years. We first verified that all chosen EOSs are viable in EMSG for the range of $\alpha$ studied: causality is maintained and $2\,M_\odot$ stars are supported. By solving the modified TOV equations, we obtained mass-radius relations, which are consistent with current NICER and GW observations. We then computed several spacetime curvature invariants: the Ricci scalar $\mathcal{R}$, Ricci contraction $\mathcal{F}$, Kretschmann scalar $\mathcal{K}$, and Weyl scalar $\mathcal{W}$. The GR case ($\alpha=0$) served as a baseline and agreed with known results. Introducing EMSG corrections ($\alpha\neq0$), we found that $\mathcal{R}$ and $\mathcal{F}$ remain largely insensitive to $\alpha$ and they vanish at the surface in all cases. By contrast, $\mathcal{K}$ and $\mathcal{W}$ exhibit significant $\alpha$-dependence, especially in the high-density core. 

The Kretschmann scalar $\mathcal{K}$ peaks at the center and decreases outward.  For the hadronic EOSs, $\mathcal{K}(r)$ is smooth; for the HQPT EOSs, the Soft and Intermediate cases show a phase transition plateau near the core (absent in the Stiff model since its central density never reaches the transition).  The effect of $\alpha$ on $\mathcal{K}$ is most evident in the core: positive $\alpha$ increases $\mathcal{K}$, while negative $\alpha$ reduces it.  The Weyl scalar $\mathcal W$ behaves oppositely: it is zero at the center and maximizes near the surface.  Crucially, $\mathcal{W}$ showed even stronger $\alpha$-dependence than $\mathcal{K}$, with a significantly broader spread of values for different $\alpha$ (especially for soft EOSs).  This suggests that the tidal curvature $\mathcal{W}$ is a more sensitive probe of strong-field deviations from GR. We find NS surface curvatures $\sim10^{14}$ times larger than the Sun’s value.  Negative $\alpha$ further increases this SC, while positive $\alpha$ decreases it.  These trends persist up to the maximum-mass configurations for the purely hadronic EOSs; for the hybrid EOSs, the phase transition introduces additional structure in the highest-mass cases, highlighting the interplay between gravity and dense-matter physics.

In summary, our results shows that EMSG can significantly modify the internal curvature of neutron stars.  In particular, the Weyl invariant $\mathcal{W}$ is extremely responsive to the EMSG coupling $\alpha$ and to EOS stiffness, making it a promising diagnostic for testing gravity in the strong-field regime.  Moreover, clear signatures of phase transitions are imprinted in the curvature profiles ($\mathcal{K}$ and $\mathcal{W}$) of the Soft and Intermediate HQPT EOSs: these features (e.g.,~the plateau in $\mathcal{K(\rho)}$ or the inflection in $\mathcal{W(\rho)}$) occur essentially independently of $\alpha$ and may provide observational imprints of exotic matter phases in NS cores.  Future improvements in observational measurements (e.g.,~higher-precision mass, radius, or potential curvature constraints from gravitational waves or lensing) could open new avenues to probe both the high-density EOS and possible modifications of gravity like EMSG.
\section{Acknowledgments}
S.G. acknowledges the National Institute of Technology Rourkela for the fellowship support. The work of S.M.~is supported by the core research grant from the Science and Engineering Research Board, a statutory body under the Department of Science and Technology, Government of India, under grant agreement number CRG/2023/007670.
\bibliographystyle{apsrev4-1}
\bibliography{main}

\end{document}